\documentclass[conference]{IEEEtran}
\IEEEoverridecommandlockouts
\usepackage{cite}
\usepackage{amsmath,amssymb,amsfonts}
\usepackage{algorithmic}
\usepackage{graphicx}
\usepackage{subcaption}
\usepackage{textcomp}
\usepackage{comment}
\usepackage{balance}

\usepackage[whole]{bxcjkjatype}

\usepackage{xcolor}
\def\BibTeX{{\rm B\kern-.05em{\sc i\kern-.025em b}\kern-.08em
    T\kern-.1667em\lower.7ex\hbox{E}\kern-.125emX}}
\begin{document}

\title{Implementation and Application of\\Multi-Format 3D Data Integration in a Cross-Device Commercial Metaverse Platform}

\author{
\IEEEauthorblockN{Masanori Ibara}
\IEEEauthorblockA{
\textit{Cluster, Inc.}\\
Tokyo, Japan \\
m.ibara@cluster.mu
}
\and
\IEEEauthorblockN{Yuichi Hiroi}
\IEEEauthorblockA{
\textit{Cluster Metaverse Lab}\\
Tokyo, Japan \\
y.hiroi@cluster.mu}
\and
\IEEEauthorblockN{Takushi Kamegai}
\IEEEauthorblockA{
\textit{Cluster, Inc.}\\
Tokyo, Japan \\
t.kamegai@cluster.mu
}
\and
\IEEEauthorblockN{Takefumi Hiraki}
\IEEEauthorblockA{
\textit{University of Tsukuba} \\
Ibaraki, Japan \\
\textit{Cluster Metaverse Lab}\\
Tokyo, Japan \\
hiraki@slis.tsukuba.ac.jp
}
}

\maketitle

\begin{abstract}
Traditionally, specialized 3D design data such as BIM and CAD have been accessible only to a select group of experts, creating significant barriers that prevent general users from participating in decision-making processes. This paper provides a systematic overview of practical insights for utilizing 3D data in industrial and architectural domains by presenting implementation cases on Cluster, a commercial cross-device metaverse platform. We analyze the characteristics and constraints of major data formats in industrial and architectural fields and organize integration workflows for metaverse environments.
Through 3D data metaverse integration cases across multiple domains, we present implementation examples of collaborative decision-making support and digital twin applications. Specifically, we demonstrate through implementation cases that multi-device access and simultaneous multi-user participation capabilities create democratic environments in industrial metaverse settings, which are difficult to achieve with conventional expert-dependent systems.


\end{abstract}


\section{Introduction}
As digital twin technology becomes more prevalent, many organizations are looking to increase their use of 3D data~\cite{Kritzinger2018-kj}. However, it remains difficult for non-technical personnel to understand BIM (Building Information Modeling) and CAD (Computer-Aided Design) data intuitively and to participate in decision-making processes in actual field operations. For example, in architectural design, most stakeholders can only review plans through drawings; in disaster prevention planning, residents can only confirm evacuation routes through floor plans; and in industrial facility operations, field workers struggle to understand design intentions. The information gap between experts and non-experts is a fundamental challenge that impedes effective collaboration and democratic decision-making.

Metaverse environments offer an effective solution to this challenge by allowing multiple users to share 3D spaces and interpret specialized data through intuitive operations. Additionally, real-time communication and collaboration foster an environment in which experts and non-experts can engage in discussions on an equal footing.

This paper provides a systematic overview of practical insights for utilizing 3D data in industrial and architectural domains, as demonstrated by implementation cases of the industrial metaverse on Cluster\footnote{https://cluster.mu/en}. With more than 35 million cumulative users, Cluster is Japan's largest metaverse platform, characterized by high accessibility through cross-device support, including smartphones, PC clients, and VR-HMDs. We analyze implementation experiences in various domains, including disaster prevention and architectural design support. We clarify processing workflows for different data formats, such as BIM, CAD, and point cloud data. We also explain approaches to solving technical challenges. Additionally, we clarify the benefits that cross-device metaverse access brings to digital twins, as well as the constraints and possibilities of current metaverse platforms. 

The primary contributions of this paper are threefold. First, we provide a systematic workflow for integrating heterogeneous 3D data formats (BIM, CAD, point cloud, and urban data) into cross-device metaverse environments. Second, we demonstrate through practical implementation cases how metaverse-based 3D data visualization can support collaborative decision-making across diverse user groups. Third, we identify technical limitations and research challenges that must be addressed for widespread adoption of industrial metaverse applications.

The remainder of this paper is organized as follows. We first review related work on 3D data integration in metaverse platforms (Sec.~\ref{sec:related_work}) and analyze the advantages of 3D data integration in cross-device metaverse environments (Sec.~\ref{sec:metaverse_benefits}). We then present a multi-format 3D data integration pipeline for commercial cross-device metaverse platforms (Sec.~\ref{sec:format_pipeline}) and demonstrate practical applications through six implementation cases (Sec.~\ref{sec:impl-cases}). Finally, we discuss technical limitations and future research challenges (Sec.~\ref{sec:limitations}).





\section{Related Work}
\label{sec:related_work}
This section reviews metaverse platforms and their capabilities for integrating specialized 3D data from industrial and architectural domains, the advantages of 3D data integration for collaborative decision-making, and digital twin implementations.



\subsection{Metaverse Platforms and Industrial Applicability}
\subsubsection{VR-Centric Platforms}
VRChat and Horizon Worlds are immersive, avatar-based communication platforms that have successfully fostered creative communities. However, they are heavily dependent on virtual reality headsets and require high-performance computers with a stable Internet connection~\cite{zhang2024should}. This makes them unsuitable for industrial applications that require the participation of diverse stakeholders, such as field workers and smartphone users~\cite{maio2023realtime}. Additionally, there are no established methods for importing BIM and CAD data while preserving embedded attribute information. This hinders the implementation of dynamic interactions and validation functions based on semantic attributes.


\subsubsection{Gaming-Centric Platforms}
Roblox, Fortnite Creative, and Minecraft are excellent platforms for user-generated content with scalable game engines. However, they lack the millimeter-level precision and attribute management capabilities required for industrial applications. In addition, accurately detecting collisions between terrain and mechanical equipment is a significant challenge~\cite{wang2021digital}.


\subsubsection{Enterprise-Focused Platforms}
NVIDIA Omniverse offers high-precision simulation and rendering quality. However, its reliance on GPU clusters and commercial licensing can create significant cost barriers for small-scale projects and mobile users. Recent reviews of metaverse applications in industrial contexts have highlighted the need for more accessible enterprise solutions~\cite{kour2025metaverse}.


\subsubsection{Specialized Platforms}
Medical education and architecture-specific platforms offer optimized user interfaces for their respective fields~\cite{tang2025evaluation, 9382657}, it requires extensive customization and deployment time, limiting democratic participation.


\subsection{3D Data Integration in Metaverse Platforms for Industrial Applications}
Various industrial sectors are exploring the integration of specialized 3D data into metaverse platforms to enable collaborative workflows. 
These integrations encompass diverse applications ranging from collaborative decision-making support to digital twin implementations. 
Sai et al. categorized digital twin applications in metaverse environments into 15 types, demonstrating the broad applicability of 3D data integration~\cite{SAI2024}. 


In design and manufacturing, Tao et al.~\cite{Tao2018DigitalTwin} and Mourtzis et al.~\cite{mourtzis2022human} proposed a digital twin framework to optimize the product design and manufacturing processes. In construction, research explores connection of BIM and VR to rapidly generate VR models of existing facilities for use in design reviews~\cite{10038293}.

In disaster prevention and emergency response training, integrating specialized 3D data into metaverse platforms enables the replication of dangerous real-world scenarios in safe virtual environments, thereby enhancing emergency response capabilities~\cite{khanal2022virtual}.
Lu et al.~\cite{LU2020102792} developed a system that uses BIM models of buildings to simulate indoor fire rescue scenarios after earthquakes in VR. Xu et al.~\cite{XU20141} constructed a VR-based firefighting training simulator with real-time evaluation capabilities.

For urban infrastructure monitoring and environmental management, Ge et al. proposed the Urban Flooding Digital Twin (UFDT) framework for urban flood management~\cite{Ge31122025}. 
UFDT demonstrates how urban 3D data integration enables multi-stakeholder collaborative intervention, and risk prediction in virtual spaces~\cite{li2021smart,white2021digital}.






\subsection{Collaborative Decision Support Through 3D Data Integration}
\label{subsec:dt_decision_support}
The integration of specialized 3D data into metaverse platforms enables new forms of collaborative decision-making previously limited to experts.
White et al.~\cite{WHITE2021103064} developed a digital model of Dublin city as an online public platform and suggested a citizen feedback system that enables residents to virtually explore urban spaces and evaluate urban planning proposals. This visualization makes it accessible to those who were previously unable to understand it and promoting participatory decision-making processes.

3D data integration for user experience design is also being explored in smart home applications. Xue et al. proposed the ``Meta-Home'' concept and discussed strategies to design digital twins of smart home environments within metaverse spaces~\cite{10.1145/3629606.3629635}.
Eneyew et al. developed systems that integrate IoT sensor data with BIM models in virtual environments~\cite{9987476}. These systems provide an integrated visualization of static structural information and dynamic environmental data, which supports facility management decision-making and represent one form of digital twin implementation.
These works demonstrate approaches that make specialized 3D data comprehensible to non-experts within metaverse environments. 

\section{Advantages of 3D Data Integration in Cross-device Metaverse Platforms}\label{sec:metaverse_benefits}

This section outlines the advantages of integrating specialized 3D data into cross-device metaverse platforms, enabling collaborative decision-making and digital twin implementations, with a focus on the industrial domain.


\subsection{Usability and Access}\label{subsec:usability_access}
Integrating specialized 3D data into metaverse environments fundamentally transforms conventional, expert-dependent approaches. The most significant innovation is operability that does not require specialized knowledge or equipment. Interfaces that enable non-experts to operate and understand specialized 3D data, such as BIM and CAD, through intuitive metaverse interactions eliminate technical barriers.

This democratization is achieved through cross-device compatibility. Access from various devices, including PCs, tablets, and smartphones, broadens user bases and eliminates the need for specialized equipment. It removes entry barriers, such as high-performance workstations and specialized software licenses required by traditional CAD systems, enabling specialized 3D data access with general consumer devices.

\subsection{Immersive, Embodied Experience}\label{subsec:immersive_experience}
One advantage of metaverse environments is the ability to interact using attribute information. Users can obtain realistic experiences within virtual spaces by assigning object-specific functions derived from 3D data attributes, such as opening and closing doors and sitting down. These functional interactions promote a dynamic spatial understanding that goes beyond viewing static 3D models.

Furthermore, through the embodied experiences of avatars, users can physically move and operate within virtual spaces, achieving experiences that closely resemble actual behavioral sensations. This type of interaction provides spatial recognition that is unattainable through conventional, screen-based visualization when it comes to understanding architectural spaces and industrial facilities.

\subsection{Collaborative Environments}\label{subsec:collaborative_env}
The essential value of metaverse technology lies in its ability to enable simultaneous, multi-user access and communication. These environments allow multiple users to examine and discuss complex 3D data simultaneously, facilitating collaborative work that was difficult to achieve with conventional, single-user CAD systems. Cross-device metaverses especially enhance real-time spatial sharing and voice communication, facilitating effective consensus building among geographically distributed stakeholders, such as indoor and outdoor users.

\subsection{Data-Driven Development Efficiency}\label{subsec:data_driven_efficiency}
From a systems development perspective, leveraging existing 3D data within standard metaverse functions minimize the need for project-specific customization. This enables rapid deployment and cost reduction through the reuse of 3D data assets and processing pipelines and efficient horizontal expansion across industrial sectors.

In addition, the metaverse provides comprehensive user behavior analytics. Unlike traditional CAD systems, metaverse platforms continuously capture detailed behavioral data, including movement patterns, interaction frequencies, areas of attention, and collaborative engagement metrics. These analyses reveal which elements of the interface cause user difficulties, where spatial navigation becomes problematic, and which collaborative features generate productive results. Development teams can use this quantitative feedback to prioritize feature improvements and reduce iterative development cycles.


\subsection{Large-Scale Data Integration and Scalability}\label{subsec:scalability}
Another important feature of metaverse environments is their ability to integrate and use large-scale urban data. By linking with public geographic information systems, such as Cesium\footnote{https://cesium.com/industries/smart-cities/} and PLATEAU\footnote{https://www.mlit.go.jp/plateau/}, it becomes possible to provide experiences on the scale of cities and infrastructure. Seamless transitions between scales, from individual buildings to entire cities, allow for a comprehensive understanding of space.


\section{Multi-Format Data Integration Pipeline}\label{sec:format_pipeline}
Effectively integrating 3D data from industrial and architectural domains into metaverse platforms is a technical challenge for enabling collaborative workflows and advanced applications such as digital twin implementations. Since these data formats have different generation processes and purposes, their advantages and constraints in metaverse integration differ significantly. This section explains the pros and cons of integrating representative 3D data from these domains into metaverse platforms, as well as processing workflows.


\subsection{BIM}\label{subsec:bim_integration}
\begin{figure}[t]
    \centering
    \includegraphics[width=1\linewidth]{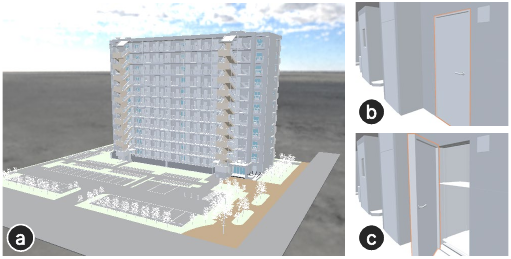}
    \caption{Integrate the BIM data into Cluster. (a) A 3D space constructed from BIM data. Multiple people can access the space simultaneously using various devices, such as VR-HMDs, PCs, and smartphones. (b, c) The door in the BIM data can be interacted (opened and closed) in Cluster. (Model by "BIM for apartment building design", Urban Renaissance Agency, Japan\protect\footnotemark)}
    \label{fig:bim_sample}
\end{figure}

\footnotetext{https://www.ur-net.go.jp/rd\_portal/UR-BIM/shugohbim.html}

BIM data is the most advantageous format for metaverse integration. Their major advantages include systematically organizing data and assigning rich attribute information, such as materials, functions, and dimensions, to architectural elements. This attribute information enables the automatic generation of interactions in metaverse environments, significantly improving development efficiency.

However, high installation costs are a constraint. BIM modeling requires specialized knowledge and time, and often BIM data cannot be used in existing projects. Furthermore, since its application scope is limited to architecture, other data formats must be used alongside BIM data for industrial machinery and equipment.



The BIM data integration workflow in Cluster is as follows (Fig.~\ref{fig:bim_sample} a):
\begin{enumerate}
    \item \textbf{Data Import}: BIM data are imported in the general-purpose Industry Foundation Classes (IFC) format. Unity engine import uses the IFC-to-glTF converter provided by ifcOpenShell\footnote{https://ifcopenshell.org/} and the UPM (Unity Package Manager) integration package distributed by the PLATEAU project. glTF is lightweight and highly versatile, making it suitable for real-time applications, such as VR/AR.
    \item \textbf{Material Replacement}: Since IFC-derived BIM models often lack appropriate materials and textures, they are automatically replaced with lightweight shaders to reduce rendering load. Materials with pre-prepared repeat textures may also be substituted based on attributes.
    \item \textbf{Texture Coordinate Generation}: In cases where meshes lack texture (UV) coordinates, world coordinates are used to generate UV coordinates. This process ensures the accurate rendering of materials with repeated textures, which significantly improves visual quality.
    \item \textbf{Interactive Element Integration}: Automatic assignment of interactive functions using attribute information is performed. For instance, door elements are automatically assigned opening/closing mechanisms (Fig.~\ref{fig:bim_sample} b, c), and chairs receive seating functions via Unity scripts. BIM shape information is analyzed to dynamically calculate pivot points for movable parts and user interaction ranges.
    \item \textbf{Rendering Load Reduction}: For large-scale building rendering, rendering load becomes problematic. Therefore, Unity's standard optimization techniques, including occlusion culling, GPU instancing, and batching, are applied to reduce rendering load.
    \item \textbf{Metaverse Environment Upload}: The final models are automatically uploaded to cloud-based metaverse environments using the metaverse software development kit (SDK). Completing this entire process on a single machine creates a seamless, reproducible pipeline.
\end{enumerate}


\subsection{CAD}\label{subsec:cad_integration}
CAD data are prevalent in many industrial domains and can be easily converted to use in metaverse environments. Since they are the standard for mechanical design and product development, there is significant potential to use existing data. This plays an important role in the implementation of industrial applications.

However, a significant constraint is the lack of attribute information. CAD data focuses on geometric shapes and often lacks semantic information, such as function and material. Technical problems also arise when converting complex shapes, such as mechanical components, into meshes. These problems include increased rendering load due to excessive mesh allocation to local areas and gap formation. Additionally, diverse formats require specialized software for conversion, except for intermediate formats such as STEP and IGES format. 

Technical difficulties arise when converting NURBS data, such as data from Rhinoceros\footnote{https://www.rhino3d.com/}, to gaming meshes due to a lack of thickness and issues with overlap. Mesh conversion from NURBS data often results in gaps or surfaces without thickness, requiring manual correction using DCC software such as Blender or Maya. The integration of interactive elements is essentially manual for each case due to the complexity of hierarchical actions and collision detection.


\begin{figure}[t]
    \centering
    \includegraphics[width=1\linewidth]{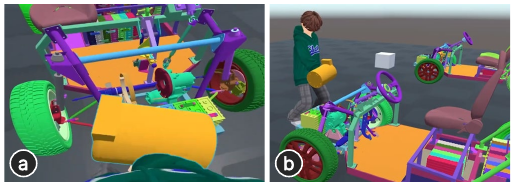}
    \caption{Assembling models using CAD data imported into Cluster. (a) First-person view. Each part is color-coded. (b) View from other users. Multiple users can collaborate to learn procedures or make design decisions. (CAD Model by OSVehicle Tabby EVO\protect\footnotemark)}
    \label{fig:cad_app}
\end{figure}
\footnotetext{https://www.openmotors.co/product/tabbyevo/}

The CAD data integration workflow in Cluster is as follows (Fig.~\ref{fig:cad_app}):
\begin{enumerate}
    \item \textbf{Data Import}: CAD data are converted to the FBX format using specialized software that minimizes the polygon count during conversion. However, NURBS and complex shape data may result in meshes with gaps or no thickness. These meshes require modification using digital content creation (DCC) software such as Blender or Maya.
    \item \textbf{Material Replacement / Texture Coordinate Generation}: Same as BIM.
    \item \textbf{Interactive Element Integration}: Due to gimmicks involving hierarchically conditioned actions and collision detection rather than simple reactive behaviors, as well as a small number of elements and settings, the process is essentially manual for each case. Independent gimmicks, such as measuring tools, are implemented as reusable components.
    \item \textbf{Rendering Load Reduction / Metaverse Environment Upload}: Same as BIM.
\end{enumerate}


\subsection{Point Cloud}\label{subsec:pointcloud_integration}

\begin{figure*}[t]
    \centering
    \includegraphics[width=1\linewidth]{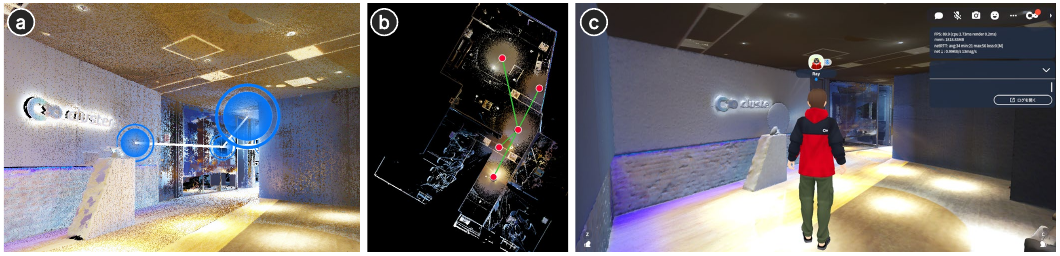}
    \caption{Point cloud data imported into Cluster. (a) Indoor point cloud data scanned with Leica BLK360 G2 and (b) its scan locations. (c) Reproduction via mesh within Cluster. Access to the space is also possible from smartphones.}
    \label{fig:point_cloud}
\end{figure*}

The advantage of point cloud data is its ability to immediately acquire data from real environments through 3D scanning technology. Direct digitization through scanning is possible in environments lacking CAD or BIM data, offering high flexibility. This is useful for facilities where renovations or updates have caused design drawings to become inconsistent with current conditions, as well as for existing structures with complex shapes.

However, significant constraints include high processing loads: generating meshes from point cloud data is computationally expensive, and data corruption is difficult to correct. In addition, the processing of complex shapes requires division, which increases labor requirements. Furthermore, high-precision scanning requires expensive equipment and specialized techniques.

The point cloud data integration workflow in Cluster is as follows (Fig.~\ref{fig:point_cloud}):
\begin{enumerate}
    \item \textbf{Point Cloud Data Acquisition}: Point cloud data are collected from real environments using 3D scanning technology.
    \item \textbf{Mesh Generation}: 3D mesh models are generated from point cloud data.
    \item \textbf{Model Correction}: Defects and missing parts in generated meshes are manually corrected.
    \item \textbf{Optimization Processing}: Reduces and optimizes polygon counts for rendering in a metaverse environment.
\end{enumerate}


\subsection{Public Urban Data}\label{subsec:urban_data}

Urban datasets, such as Cesium and PLATEAU, enable comprehensive 3D environments on an urban scale. These datasets support the implementation of digital twins in broad regional contexts that extend beyond individual buildings. As public data maintained by government agencies, these datasets are highly reliable and continuously updated.

However, the level of detail varies by region. Depending on the region, the quality can vary greatly, ranging from buildings represented as simple white boxes to detailed reproductions that include guardrails and monuments. In addition, technical challenges arise during data integration. For example, discrepancies may arise between road and terrain information, and building heights may not align with terrain models. Due to the large volume of urban-scale data, it is essential to optimize rendering loads, including reducing the resolution of terrain models.

\begin{figure}[t]
    \centering
    \includegraphics[width=1\linewidth]{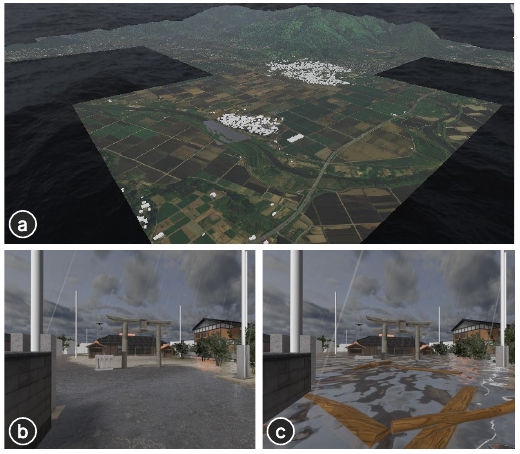}
    \caption{Flood simulation of a city reproduced on Cluster. (a) 3D data of the city integrated from PLATEAU. Flooding was simulated according to elevation. (b, c) Flooding as seen from the perspective of real humans.}
    \label{fig:plateau_app}
\end{figure}

The process of integrating city data into Cluster is described below using Japan's official city data from PLATEAU as an example (Fig.~\ref{fig:plateau_app} a).
\begin{enumerate}
\item \textbf{Range Selection and Import}: Using the PLATEAU project SDK, we carefully select the necessary ranges and import the data. First, we confirm that the quality of the data matches the intended use.
\item \textbf{Data Correction}: Manual corrections are made for discrepancies between road and terrain information, as well as for cases where the heights of the building group do not match the terrain models.
\item \textbf{Reduction Processing}: Unnecessary objects are deleted. In particular, buildings that are invisible from overhead viewpoints or hidden by terrain are automatically deleted using a dedicated EditorScript that determines visibility through raycasting.
\end{enumerate}

\section{Implementation Cases}\label{sec:impl-cases}
This section presents case studies on the implementation of 3D data integration and metaverse-based applications on Cluster. Some cases demonstrate digital twin functionality through real-time sensor integration, while others focus on collaborative 3D data utilization. Each study demonstrates how to apply the advantages of metaverse technology and integrate processing techniques for various data formats, as discussed in Sec.~\ref{sec:metaverse_benefits} and Sec.~\ref{sec:format_pipeline}, respectively, in each application domain.

The following case studies present implementation examples along with technical challenges and preliminary user feedback. While these observations provide valuable insights into practical deployment, systematic user evaluation remains future work.


\begin{figure*}[t]
    \centering
    \includegraphics[width=1\linewidth]{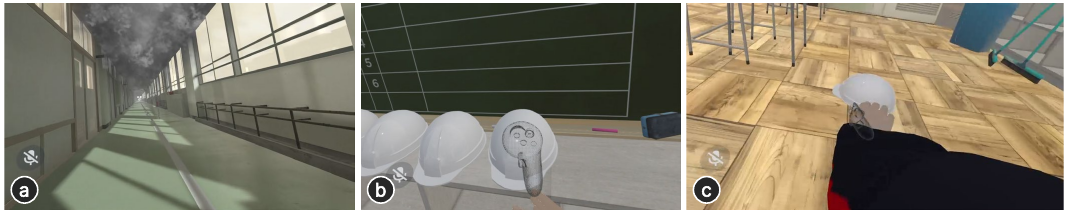}
    \caption{Simulation of evacuation drills using school building data in Cluster. (a) Multiple people can confirm smoke flow to create a more realistic situation. (b, c) Participants can experience physical actions such as going to the location where helmets are stored and actually holding them in their hands.}
    \label{fig:school_app}
\end{figure*}

\subsection{Disaster Experience}\label{subsec:app-disaster}
In Cluster, we use PLATEAU to recreate the terrain and buildings and render the nearby structures in high resolution, allowing us to visualize the arrival of a flood on a river in real time (Fig.~\ref{fig:plateau_app}). This allows participants to experience damage scenarios from realistic perspectives during disaster prevention events. For earthquake, fire and flood disaster evacuation simulations, participants could verify actual evacuation routes while experiencing appropriate levels of fear and situational awareness in virtual space.

This case study combined PLATEAU data with FBX assets to utilize a broad geographical and detailed architectural information. This approach provides realistic and immersive experience unavailable in conventional 2D disaster simulations. It effectively improved the awareness of disaster prevention by participants (Sec.~\ref{subsec:scalability}). The access of smartphones allowed ordinary citizens to participate without specialized equipment, contributing to the democratization of disaster prevention education (Sec.~\ref{subsec:usability_access}).

Implementation challenge involved reproducing complex urban water flow dynamics while maintaining real-time interaction within cross-device rendering constraints.
Preliminary participant feedback confirmed effectiveness of realistic and fear-inducing VR experiences of residential areas, and improved evacuation awareness, particularly for height-adjusted child perspectives.


\subsection{Evacuation Training and Route Optimization}\label{subsec:app-evacuate}
We created evacuation training simulations based on detailed architectural information using BIM data integration (Sec.~\ref{subsec:bim_integration}). We simulate a fire scenario in schools, reproducing the entire processes, from carrying helmets to reaching outdoor safety areas (Fig.~\ref{fig:school_app}). In the metaverse, we track evacuation behavior and collect movement and gaze direction information from each user to visualize route bottlenecks. These high-frequency human-based simulations in a virtual space enable us to optimize evacuation routes without repeated physical drills.

This approach enables safe implementation of large-scale evacuation simulations and the reproduction of dangerous situations that would be challenging to recreate in physical training. Using collaborative environments (Sec.~\ref{subsec:collaborative_env}), we statistically analyze (Sec.~\ref{subsec:data_driven_efficiency}) the collected data and apply it to improving facility designs and optimizing evacuation plans. 
Collected data enables statistical analysis for facility design improvement and evacuation plan optimization.

Implementation faced technical challenges in debugging difficulties arising from complex user movement paths and accurate reproduction of object physical behavior during earthquake scenarios. Preliminary user feedback appreciated simulation realism of earthquake and smoke, while noting VR motion sickness concerns.



\subsection{Design Decision Support}\label{subsec:app-bim-design}
We integrated BIM data into Cluster and assigned interactions based on BIM attribute information, such as opening and closing (Sec.~\ref{subsec:bim_integration}, Fig.~\ref{fig:bim_sample} b, c). This creates environment where general stakeholders and clients without specialized knowledge can confirm and discuss design proposals while observing actual behaviors.

This setup enables low-cost, high-frequency presentations of one-off manufactured products, such as ships and buildings, for decision-makers, even in the event of design changes. Adding realistic measurement tools, such as tape measures, offers advantages in metaverse construction. These tools allow confirmation of interior details and the convenience and feasibility of movement routes for multiple people. Through immersive interactions and embodied experiences (Sec.~\ref{subsec:immersive_experience}), we facilitate spatial recognition, an intuitive experience that was difficult to achieve through conventional drawing-based confirmation. This promotes quality improvement of the design and the formation of consensus among stakeholders. Real-time collaborative review sessions enable efficient decision-making among geographically distributed stakeholders (Sec.~\ref{subsec:collaborative_env}).

Technical constraints included processing times for large-volume BIM data conversion and optimization processing. Preliminary participant feedback demonstrated high appreciation for smartphone accessibility, confirming the feasibility of democratic design participation environments that eliminate specialized equipment requirements.



\subsection{Work Simulation Using CAD}\label{subsec:app-cad-design}
We integrated CAD data into Cluster (Sec.~\ref{subsec:cad_integration}, Fig.~\ref{fig:cad_app}) and assigned interactions that reproduce assembly, disassembly, and processing operations. Contact and dangerous actions are communicated safely to users via visual warnings. Realistic training sessions with multiple participants  are also possible.

Using actual equipment for industrial training is costly and risky. However, metaverse environments enable repeated practice through embodied experiences (Sec.~\ref{subsec:immersive_experience}) and the safe simulation of dangerous situations. 
Visualizing work procedures, having learners and instructors collaborate in the same space (Sec.~\ref{subsec:collaborative_env}), and providing feedback based on behavioral data analysis greatly improve understanding and training efficiency (Sec.~\ref{subsec:data_driven_efficiency}).

Implementation challenges centered on processing objects without thickness and increased manual processing overhead for combining meshes separated by individual surfaces. Preliminary participant feedback confirmed intuitive disassembly operation and effective full-scale visualization, promoting understanding of complex mechanical structures.



\begin{figure}[t]
    \centering
    \includegraphics[width=1\linewidth]{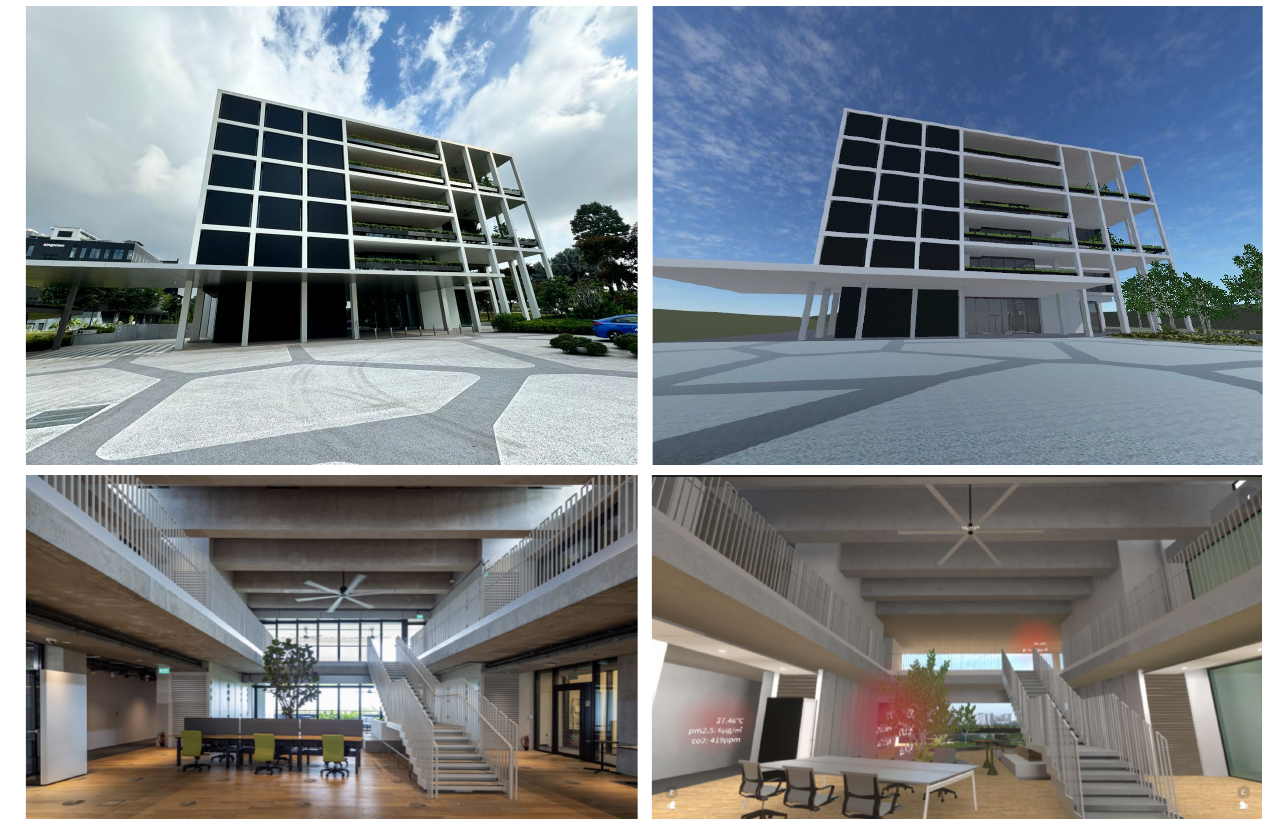}
    \caption{Our digital twin demonstration of the smart building on Cluster~\cite{10973013} (Left) the actual building interior, (Right) a virtual reproduction of the same space. As shown in the lower-right figure, part of the space is colored red, and information from IoT sensors, such as temperature and CO$_2$ concentration, is visualized.}
    \label{fig:kajima_digital_twin}
    \vspace{-3mm}
\end{figure}

\subsection{Environmental Monitoring through IoT Sensor Integration}\label{subsec:app-iot-integrate}
We linked the BIM data of a smart office building (Sec.~\ref{subsec:bim_integration}) with IoT sensor networks to create a metaverse environment that reflects real-time environmental data, such as temperature, humidity, CO$_2$ concentration, and wind speed~\cite{10973013} (Fig.~\ref{fig:kajima_digital_twin}). 
Changes in physical environments are synchronized with the virtual environment with approximately 0.3 seconds latency. This allows users to easily understand environmental conditions through color-coded temperature displays, numerical air quality data, and wind speed via sound effects.

Unlike conventional environmental monitoring systems, metaverse environments allow multiple users to share environmental information simultaneously in 3D space and conduct real-time consultations and decision-making processes (Sec.~\ref{subsec:collaborative_env}). Remote users can experience the same environmental conditions as those in physical facilities, facilitating efficient facility management and collaborative discussions about energy-saving strategies. Multi-device support enables participation in environmental monitoring via smartphones, thus contributing to the democratization of facility management (Sec.~\ref{subsec:usability_access}).

The technical challenges included the complexity of building relay servers to connect the Cumulocity API, the industrial IoT platform, and the external communication APIs of Cluster. Preliminary user feedback appreciated visual comprehensibility, while requesting more comprehensive spatial navigation functionality.



\subsection{Equipment Transport Simulation}\label{subsec:app-equipment}
Cluster integration of point cloud data (Sec.~\ref{subsec:pointcloud_integration}) addresses equipment transport planning by reproducing detailed 3D models of existing facilities. Even in environments where frequent equipment updates prevent drawings from remaining current, scanning allows verification of construction environments with minimal effort (Sec.~\ref{subsec:data_driven_efficiency}). Life-size perspectives enable visualization of the actual work environment and identification of areas requiring attention or posing danger in advance (Sec.~\ref{subsec:immersive_experience}). Once constructed, multiple people can participate in the virtual space to practice transport procedures, thereby improving construction efficiency and reducing risk (Sec.~\ref{subsec:collaborative_env}).

Implementation faced technical constraints including acquiring uniform high-density point cloud data, time costs for post-processing involving unnecessary data removal and mesh conversion, and inability to scan transparent materials such as glass. Preliminary participant feedback appreciated spatial reproduction realism, while noting concerns about visible holes from data deficiencies, requiring point cloud data quality improvement and post-processing enhancement.

\section{Technical Limitations and Challenges}\label{sec:limitations}
This section analyzes technical constraints revealed through implementation cases, implementation challenges for future integration of industrial 3D data into cross-device metaverse platforms, and academic research challenges.

\subsection{Analysis of Technical Constraints}
Constraints on cross-device compatibility create a dilemma between democratization and advanced functionality. Although supporting devices ranging from smartphones to virtual headsets eliminates participation barriers, it necessitates system designs limited by lowest-performing devices. In disaster experience implementations featuring complex water flow dynamics (Sec.~\ref{subsec:app-disaster}) and evacuation training with physics simulation (Sec.~\ref{subsec:app-evacuate}), maintaining real-time processing on low-performance devices, creating an unavoidable trade-off between visual quality and processing speed.

Semantic gaps in heterogeneous data integration represent obstacle for automation. Large-volume BIM data conversion, correction of CAD objects lacking thickness, and point cloud post-processing create format-specific challenges that prevent automated pipeline construction. BIM, CAD, point clouds, and urban data differ in technical format, design philosophy, and intended use, making integration processing interpretive work.

Large-scale data processing with real-time performance introduces scalability limitations. Regional quality variations in PLATEAU urban data (Sec.~\ref{subsec:app-disaster}), BIM data rendering loads (Sec.~\ref{subsec:app-bim-design}), and high-density point cloud processing (Sec.~\ref{subsec:app-equipment}) exceeds the capacity of single processing pipelines. These challenges highlight balancing standardized processing with individualized optimization in industrial metaverse deployment.

\subsection{Required Platform Extensions}
Recent developments in high-precision rendering, such as Neural Radiance Fields (NeRF)~\cite{mildenhall2020nerf} and 3D Gaussian Splatting~\cite{kerbl3Dgaussians}, offer solutions to point cloud constraints. These continuous 3D representation technologies naturally reproduce transparent materials, such as glass, and automatically repair holes through data interpolation. This enables high-quality rendering under cross-device constraints~\cite{duckworth2023smerf}.

Cloud-based, distributed rendering addresses cross-device constraints~\cite{11003161}. Device-adaptive quality control enables dynamic delivery of lightweight versions for smartphones and high-definition versions for VR-HMDs, allowing simultaneous experiences at different quality levels. Edge computing integration also enables streaming high-quality rendering while minimizing network latency~\cite{9712215}.

Large Language Models (LLMs) specialized in architecture and industry domains address semantic gaps in heterogeneous data integration~\cite{10.1145/3706598.3714224, 9974380}. These systems interpret BIM architectural attributes and CAD mechanical elements to automatic completion of attribute information, generation of interaction elements, and detection and correction of quality defects~\cite{Kurai2025-bf, 10765209}.

Machine learning-based adaptive Level of Detail (LOD)~\cite{windisch2025lod, 10.1145/3641519.3657420} systems promise improvements in large-scale data processing. By dynamically analyzing user attention points, movement patterns, and device performance to stream only necessary detail levels in real-time, these systems enable compatibility between city-scale data and high-definition individual building display, offering fundamental solutions to scalability problems.

\subsection{Future Research Challenges}
Elucidating the mechanisms by which collective intelligence emerges in virtual spaces is the most critical academic challenge. Decision-making processes involving diverse stakeholders in 3D spaces exhibit cognitive and social dynamics that are qualitatively different from those in conventional 2D meeting systems~\cite{10.1145/3567741}. It is essential to investigate the quantitative effects of spatial arrangement, visual sharing, and embodied experience.

Data consistency and conflict resolution in real-time collaborative environments presents technical challenges~\cite{10.1145/3544548.3581136}. State synchronization and history management for multiple users simultaneously manipulate 3D data involves spatial and temporal complexities beyond conventional database technologies.

Quality assurance and traceability mechanisms are prerequisites for industrial applications~\cite{10.1145/3334480.3375027}. When applying design decisions and training results from metaverse environments to physical systems, the systems must verify the correlations between actions in the virtual environment and outcomes in the real world, prevent the propagation of errors, and clarify responsibility attribution. These systems are foundational technologies that ensure the reliability and practicality of industrial metaverse applications.

\section{Conclusion}
This paper introduced a practical framework for using 3D data in the industrial and architectural sectors by creating an industrial metaverse with Cluster, cross-device metaverse platform. The integration workflow was systematized for BIM, CAD, point clouds, and PLATEAU urban data. This process revealed optimization methods corresponding to the characteristics of each data format.
Implementation cases demonstrated that combining metaverses and digital twins can create environments that enable democratic participation and surpass conventional, expert-dependent systems. Eliminating participation barriers through access between devices and supporting collaborative decision-making are important milestones in the social implementation of the industrial metaverse.

This position paper provides practitioners with practical guidelines for using 3D data in metaverse environments and establishes an academic foundation for developing industrial metaverses. It paves the way for new collaborative environments that integrate physical and virtual spaces.







\section*{Acknowledgment}
This study was supported by JST Moonshot Research \& Development Program Grant Number JPMJMS2013 and JST ASPIRE Grant Number JPMJAP2327, Japan.

\balance
\bibliographystyle{IEEEtran}
\bibliography{template}


\end{document}